\documentclass[oldversion]{aa}
\usepackage{graphicx}
\usepackage{txfonts}
\usepackage{natbib}
\usepackage{subfigure}

\bibpunct{(}{)}{;}{a}{}{,}

\begin{document}


   \title{Formation of cold filaments in cooling flow clusters}

   \author{Y. Revaz\inst{1} 
          \and 
	  F. Combes\inst{1}
	  \and 
	  P. Salom\'e\inst{2}}

   \offprints{Y. Revaz}

   \institute{LERMA, Observatoire de Paris, 61 av. de l'Observatoire, 75014 Paris, France
              \and
	      Institut of Radio Astronomy (IRAM), Domaine Universitaire, 300, rue de la piscine, F-38400 St-Martin d'H\`eres
              }

   \date{Received -- -- 20--/ Accepted -- -- 20--}

 
  \abstract
   {
   Emission-lines in the form of filamentary structures is common in bright clusters characterized by short cooling times.
   In the Perseus cluster, cold molecular gas, tightly linked to the H$\alpha$ filaments, has been recently revealed by CO observations.
   In order to understand the origin of these filamentary structures, 
   we have investigated the evolution of the hot ICM gas perturbed by the AGN central activity in a Perseus like cluster.
   Using very-high resolution TreeSPH simulations combined with a multiphase model and a model of plasma bubbles,
   we have been able to follow the density and temperature evolution of the disturbed ICM gas around the bubbles.
   Our simulations show that a fraction of the $1-2\,\rm{keV}$ gas present at the center of clusters is trapped and entrained  
   by the rising buoyant bubble to higher radius where the AGN heating is less efficient. 
   The radiative cooling makes it cool in a few tens of Myr below $10^4\,\rm{K}$, forming cold filamentary structures 
   in the wake and in the rim of the bubbles.
   %
   The predicted cold gas formed outside the cluster center is in agreement with the total CO mass and density profile of the observed molecular gas
   as well as with the kinematics of the H$\alpha$ filaments.
   This scenario explains the H$\alpha$ and CO filaments observed in luminous clusters
   without contradicting the observed lack of $1\,\rm{keV}$ gas.
   It also emphasizes that if the AGN feedback provides some heating (negative feedback) it also 
   perturbs the ICM, increasing its cooling (positive feedback).
   }


   \maketitle

%

\section{Introduction}

Cooling flows in clusters of galaxies has been a puzzling topic for more than 30 years \citep{fabian94,peterson06}. 
While early models of X-ray clusters \citep{cowie77} applied to observations predict a large cold gas mass 
accretion rate, the recent data from Chandra and XMM-Newton ruled out such large amounts of cooling \citep{peterson03}. 
There are now numerous observations showing that the AGN activity may be a
clue to the cooling flow problem \citep{fabian00,mcnamara00,owen00,mcnamara00,fabian03b,fabian03,blanton03,wise06,fabian06,mcnamara07}.
The emerging picture is that cooling flows are considerably reduced by the heating provided by the
AGN feedback \citep{bruggen01,churazov01,quilis01,bohringer02,reynolds02,omma04b,omma04a,sijacki06a,heinz06,cattaneo07}.
The bulk of the ICM gas remains hot ($\ge 2\,\rm{keV}$) down to the cluster center, where it is accreted
by the central black hole (Bondi accretion). 

However, in this picture, it is difficult to understand the origin of 
the observed cold neutral gas \citep{edge01,edge03,salome03,salome04,salome06} and the filamentary H$\alpha$ 
structures \citep{conselice01,crawford05}.
The connection between radio source and cold CO gas \citep{salome04} as well as the excess of optical/UV/IR light
in clusters with central cooling time less than $3\,\rm{Gyr}$ are signs of the link between the filaments, the cooling and 
the AGN activity.
Very deep Chandra observation of the Perseus cluster shows a clear correlation between the soft X-ray emission 
(between $0.5$ and $1\,\rm{keV}$) and the optical filamentary H$\alpha$ structures \citep{fabian03,fabian06}.
As the filaments cover a large range of cooling time within the hot gas, it is unlikely that they 
form by direct cooling of the $2\,\rm{keV}$ component. Moreover, the kinematics of the filament in H$\alpha$ 
\citep{hatch06} and soft X-rays \citep{fabian06} matches the gas flow under a buoyantly rising bubble.
This led \citet{fabian03,fabian06} and \citet{hatch06} to propose that the filaments have been pulled 
out from a central reservoir of cold and warm gas by the action of radio plasma bubbles inflated by AGN jets.

In this letter, without answering the question of the inner quenching of the cooling flow, we present a series of very high 
resolution TreeSPH simulations of a Perseus type cluster including AGN feedback. 
We show that buoyantly rising bubbles entrain a fraction of the $1-2\,\rm{keV}$ central ICM gas 
to much higher radius where it has time to cool and fall back to the cluster center, forming cold filamentary 
structures ($T < 10^4$).




\section{The model}\label{model}


\subsection{Initial mass distribution}

We have constructed a cluster model in order to fit the Perseus data from \citet{sanders04}.
The initial conditions are based on two components, a dark halo and a hot gas phase. 
Both share the same $\beta$-profile with $\beta=2/3$ (pseudo-isothermal sphere), with a core radius of 
$40\,\rm{kpc}$ and a central mass density of $\rho_0 = 1.38\times 10^{7}\,\rm{M_\odot  kpc^{-3}}$.
The gas represents $15\%$ of the total mass and is treated as self-gravitating in the fixed potential of the halo.
Assuming a hydrogen mass fraction of 0.76, we found a central electron density of $0.073\,\rm{e^{-}/cm^3}$
and a temperature of $2.85\times 10^8\,\rm{K}$. 
The model is truncated at a radius of $2\,\rm{Mpc}$, giving a total mass of $5.3\times 10^{14}\,\rm{M_\odot}$.
With these parameters, the mean cooling time at the center is about $500\,\rm{Myr}$.

\subsection{Gas physics}\label{physics}

Computing the interaction between the AGN and the hot ICM requires us to deal with a huge range of
temperatures, spanning from more than $10^8\,\rm{K}$ down to $10^4\,\rm{K}$. This may be achieved
by assuming a multiphase behavior of the gas where hot and cold gas represents two distinct phases.

The warm-hot gas phase ($T>10^4\,\rm{K}$) is modeled by an ideal gas with an adiabatic index of $5/3$.
The radiative cooling of the gas is computed from the normalized cooling function tabulated by 
\citet{sutherland93}, assuming an abundance of one third solar \citep{sanders04}.
Hydrodynamical equations are integrated using the Lagrangian SPH technique.

Below $10^4\,\rm{K}$, instead of quenching the cooling, the gas is assumed to change phase, becoming
very cold and clumpy, as commonly observed in the ISM. In this phase, the semi-collisional nature of the
gas is well modeled by the sticky particle technique widely used to compute
the behavior of the ISM in the disk of galaxies \citep{schwarz81,combes85,semelin02,bournaud02}.

We take into account the possibility of forming stars from the cold phase. This is computed following
the standard process described in \citet{katz96} which simulates quite well a Schmidt law.
We also take into account supernova feedback by injecting energy in both thermal ($10\%$) and kinetic ($90\%$) form,
assuming that $10^{48}\,\rm{erg}$ are released for each solar mass formed.


\subsection{AGN feedback}\label{feedback}

We have modeled the AGN feedback by instantaneously generating bubbles of gas in the ICM
This technique, which has already been used by several authors
\citep{bruggen01,churazov01,saxton01,sijacki06a,sijacki07,gardini07}, 
does not consider the evolution of the jet at its origin that may be modeled by mass or energy injection 
\citep{reynolds02,quilis01,bruggen02a,bruggen02b,bruggen03,dallavecchia04,omma04a,omma04b,heinz06,cattaneo07}.
We have implemented a new bubble generator allowing us to control precisely the relative pressure, density and
specific energy of the bubble with respect to the ambient gas. We have also taken into account the presence
of the puzzling cooler and denser gas forming a bright shell around the radio bubbles 
\citep{fabian00,mcnamara00,blanton01,nulsen02,blanton03,fabian03,fabian06,mcnamara07}, asssuming that it is
the result of ambient gas pushed out by the inflating bubble. 
From these assumptions, we can generate bubbles with densities differing from the ambient gas,
without violating the total mass conservation. 
In a given sphere of radius $R_{\rm{B}}$ at a distance $D_{\rm{B}}$ from the cluster center,
the density $\rho_i$, mass $m_i$, specific energy $u_i$ and pressure $P_i$ of a particle $i$ are modified as follows~:
%
%
%
	\begin{eqnarray}
        \rho_i^{\rm{ICM}}	\longrightarrow 	\rho_i^{\rm{B}} = \delta(R)\cdot \rho_i^{\rm{ICM}}, &&
	m_i^{\rm{ICM}}		\longrightarrow 	m_i^{\rm{B}} = \delta(R) \cdot m_i^{\rm{ICM}}, \nonumber\\
	u_i^{\rm{ICM}}		\longrightarrow         u_i^{\rm{B}} = \frac{\alpha(R)}{\delta(R)}\cdot u_i^{\rm{ICM}}, &&
	P_i^{\rm{ICM}}		\longrightarrow 	P_i^{\rm{B}} = \alpha(R) \cdot P_i^{\rm{ICM}},\nonumber\\
	\label{mup}
	\end{eqnarray}
where
	\begin{equation}
	\delta(R) = \left\{ \begin{array}{ll}
			 \delta_0							& \textrm{if $R<R_{\rm{B}}$}\\
			 \delta_0' \exp \left(\frac{R_{\rm{B}}-R}{R_{\rm{c}}}\right)+1	& \textrm{if $R \ge R_{\rm{B}}$}
	                 \end{array} \right.,
	\end{equation}
	\begin{equation}
	\alpha(R) = \left\{ \begin{array}{ll}
			 \alpha_0								& \textrm{if $R<R_{\rm{B}}$}\\
			 (\alpha_0-1) \exp \left(\frac{R_{\rm{B}}-R}{R_{\rm{c}}}\right)+1	& \textrm{if $R \ge R_{\rm{B}}$}
	                 \end{array} \right.,
	\end{equation}
where the radius $R$ is computed with respect to the bubble center.
The thickness of the rim is parametrized by $R_{\rm{c}}$, set to $1/4$ of the bubble radius. 
The density and pressure of the bubble with respect of the ambiant cluster gas are controlled by parameters $\delta_0$
and $\alpha_0$ (see Table~\ref{params}).
When $\alpha_0$ is set to 1, the pressure of the bubble is not modified with respect to the ICM
(see Eq.~\ref{mup}). In this case, from first thermodynamic principles, no energy is injected into the system
\footnote{In fact the total energy is slightly modified, due to the particle mass and density changes which imply a change in the total 
potential. However, this change is very small compared to the total potential of the cluster and may be neglected.}.
The constant $\delta_0'$ is computed in order to conserve the total mass.
According to observations, the bubble rims are slightly denser and cooler than the ambient gas, while the pressure remains unchanged.

\subsection{Simulations and parameters}

The models have been run using a modified version of the 
parallel TreeSPH code Gadget2 \citep{springel05} which conserves 
both energy and entropy. Additional physics have been included as described in paragraph~\ref{physics} and \ref{feedback}.

In order to reach a very high resolution in the cluster central region (up to $200\,\rm{kpc}$), 
we have constructed our model using a multi-resolution technique. The models are made out of 7 cells where
two consecutive cells differ by their resolution only by a factor of two, ensuring
the stability of the integration of the motion equations. This technique allows us to compute the evolution
of the bubbles within a central region of $200\,\rm{kpc}$, reaching a mass resolution of $2.4\,\times 10^6\,\rm{M_\odot}$
and a spatial resolution of about $1\,\rm{kpc}$, appreciably better than previous works using similar Lagrangian codes. 
The total number of particles is 6'041'002, equivalent to 33'554'432 particles
in a similar model where no multi-resolution technique is used. 

In order to explore the effect of the bubbles on the ambient gas, we have run several simulations exploring all the parameters. 
We have completed our study with a control run where no bubble is launched. This simulation is described in Appendix~\ref{appendix1}. 
Hereafter, we focus on a subset of 5 representative simulations.
For each simulation, the parameters are listed in Table~\ref{params}. Bubbles are created at $T=20\,\rm{Myr}$ and are followed 
for $600\,\rm{Myr}$.

Except for model 1, we have set $\alpha_0=1$ (bubbles are in pressure equilibrium with the ambient gas). 
This choice is motivated by the work of \citet{cattaneo07} showing that the bulk of the AGN feedback energy
can be deposited at the center of the cluster in the form of thermal energy. Consequently, as we have not implemented
central thermal feedback, our models are subject to the cooling catastrophe, occurring in the $10$ central kpc.
%
%
\begin{table}
    \begin{tabular}{c c c c c c c}
    \hline
    \hline
    Id  &  $D_{\rm{B}}\,[\rm{kpc}]$ & $R_{\rm{B}}\,[\rm{kpc}]$ & $\alpha_0$ & $\delta_0$ & $E_{\rm{B}}\,[10^{60}\,\rm{erg}]$ & $T_{\rm{B}} [10^7\,\rm{K}]$\\
    \hline
    \hline
    1	&  50 & 30 & 1.4 & 0.5 &  0.8 & 4.7\\	
    2	&  50 &	30 & 1.0 & 0.5 &  0.0 & 4.1\\	
    3	&  50 &	30 & 1.0 & 0.1 &  0.0 & 19.0\\	
    4	&  30 &	10 & 1.0 & 0.5 &  0.0 &  4.5\\	
    5	&  10 &	 8 & 1.0 & 0.1 &  0.0 & 30.2\\	
    \hline
    \end{tabular}
    \caption[]{List of simulations and respective parameters. See Section~\ref{model} for the meaning of each parameter.}
    \label{params}
\end{table}
%


\section{Results}\label{results}


\subsection{Global evolution of the bubbles}

\begin{figure}
  \resizebox{\hsize}{!}{\includegraphics[angle=0]{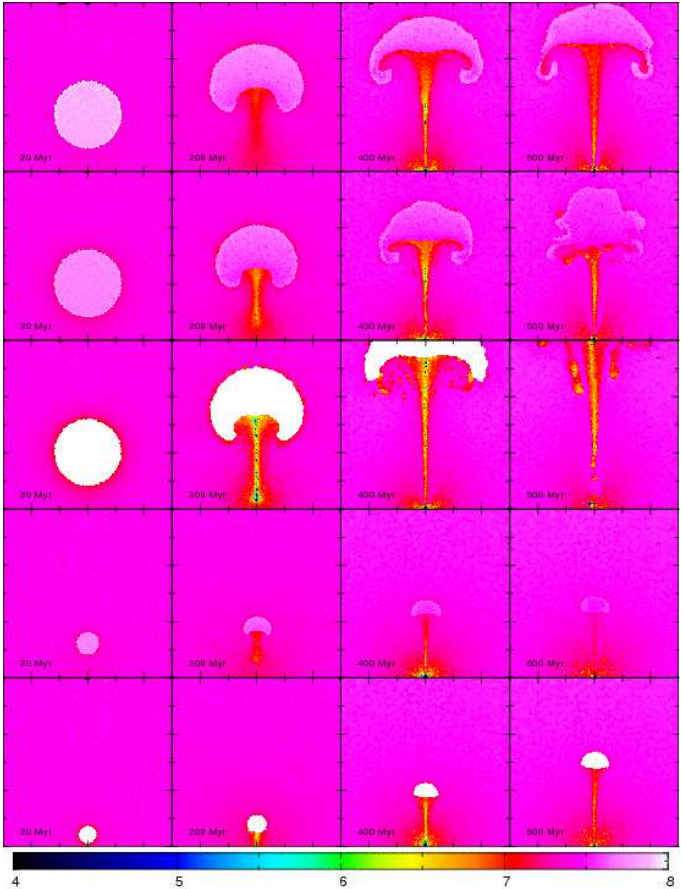}}
  \caption{Temperature map of the evolution of bubbles for models 1 (top) to 5 (bottom line). Colors represent
  the logarithm of the temperature between $10^4$ and $10^8\,\rm{K}$ as indicated by the color bar.
  Cold gas is represented by blue colors. The temperatures are computed from a thin slice of $2\,\rm{kpc}$
  width centered on $y=0$.
  The width and height of each box is $150\,\rm{kpc}$. Time increases from left to right.}
  \label{bubbles}
  \end{figure}

The evolution of the bubbles for the 5 models is displayed in Fig.~\ref{bubbles}. 
The first stage is very similar to previous
works. Once generated, the bubbles quickly start moving upwards, reaching velocities of between
$50$ and $300\,\rm{km/s}$ and forming typical atomic mushrooms. 
However, contrary to Eulerian simulations, the latter phase does not show the formation of a torus.
The bubble is only weakly affected by hydrodynamic instabilities (see \citet{agertz06} for a complete
discussion of the damping of Kelvin-Helmoltz and Rayleigh-Taylor instabilities in SPH codes) and dissolves 
only when reaching regions of similar specific entropy.
%

\subsection{Cold gas formation}\label{cold_trunk}


A common feature between the 5 simulations is the systematic formation of very cold gas in the wake of the bubbles,
far from the cluster core\footnote{See Appendix~\ref{appendix1} for an estimation of the cooling 
core in a control simulation run without feedback}.
In a few Myrs, a substantial fraction of gas cools from the ambient ICM temperature ($1-2\rm{keV}$) 
down to $10^6\,\rm{K}$, with a small fraction of it reaching temperatures below $10^4\,\rm{K}$ and forming a thin 
trunk (see Fig.~\ref{bubbles}). 
%
The presence of cooler gas in the trunk of a buoyantly rising bubble already has been mentioned by
different authors \citep{churazov01,bruggen01,saxton01,gardini07}. 
However, in all these simulations, the gas temperature never falls below $5\times 10^6\,\rm{K}$.
The high resolution obtained in our simulations reinforced by the Lagrangian technique enables us to resolve
higher densities, critical for the correct estimation of the cooling time.

\begin{figure}
  \resizebox{\hsize}{!}{\includegraphics[angle=-90]{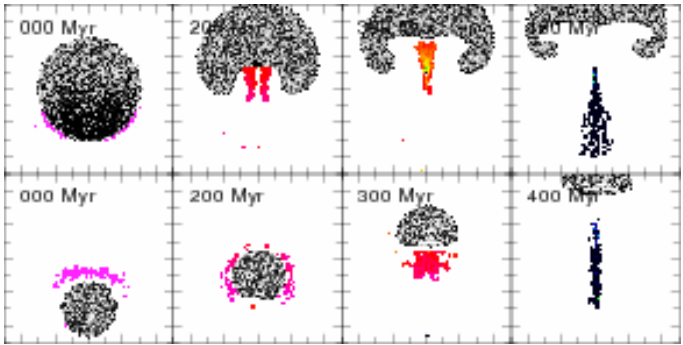}}
  \caption{Evolution of the gas around the bubble of model 2 (top) and 5 (bottom) that forms the cold filament ($T<10^6\,\rm{K}$)
  at $t=400\,\rm{Myr}$. Colors have the same meaning as in Fig.~\ref{bubbles}.
  The plasma bubble is represented in grey. The box size is respectively $50\times50$ (top) 
  and $25\times25\,\rm{kpc}$(bottom).}
  \label{bubble19_24_400}
  \end{figure}

In our simulations, the filaments are formed by $1-2\,\rm{keV}$ gas (observed in the central regions of
the Perseus cluster \citep{fabian06}) having short cooling time (between $400$ and $800\,\rm{Myr}$) at the time the bubble is formed. 
In the case of large bubbles (models 1,2,3) this gas is situated below the bubble. 
When smaller bubbles are created near the center (models 4,5), gas with sufficiently short cooling times is also found above the bubble.
In both cases, this gas is trapped by the rising bubble and entrained outside the cooling cluster core where 
its sufficiently short cooling time allows it to cool quickly to very low temperatures. 
Fig.~\ref{bubble19_24_400} shows the evolution of particles becoming part of the filaments 
($T<10^6\,\rm{K}$) at $t=400\,\rm{Myr}$ for model 2 and 5 (See also Fig.~\ref{bubble19_600} and Fig.~\ref{bubble24_600} for the
evolution of particles becoming part of the filaments at $t=600\,\rm{Myr}$).
At the beginning of simulation 2, particles being in the filaments at $t=400\,\rm{Myr}$ (resp. $t=600\,\rm{Myr}$) are part of the bottom rim of the bubble at 
a distance of between $20$ and $40\,\rm{kpc}$ (resp. $30$ and $50\,\rm{kpc}$) of the center, with a temperature between $1$ and $2\,\rm{keV}$.
During the rising of the bubble, this gas is dragged up to $80\,\rm{kpc}$  (resp. $90\,\rm{kpc}$). 
The radiative cooling slowly decreases the gas temperature, which reaches $1\,\rm{keV}$ around $t=300\,\rm{Myr}$.
At these temperatures, the gas is no longer supported by the pressure and falls towards the center while being compressed.
Its density is enhanced by a factor of 10 which substantially decreases its cooling time
\footnote{At these temperatures and densities, the cooling is so strong that it dominates the adiabatic compression heating.}. 
A very cold thin filament with a temperature below $10^4\,\rm{K}$ is then formed. 
The main difference of model 5 is that the bubble is formed at smaller radius. 
Gas with a short enough cooling time ($<600\,\rm{Myr}$) is initially situated above the bubble.
As for the model 2, it is dragged at higher altitude (up to $60\,\rm{kpc}$) and crossed by the bubble, before reaching 
low temperatures and forming the cold observed filament.


%
%

%

\subsection{Comparison with observations}\label{comparison}

While a more complete model, including thermal heat, visosity and turbulence will be needed in the future,
in Appendix~\ref{appendix2} we show that the CO gas observed in the Perseus cluster may be compatible with 
the production of cold gas by a dozen of bubble pairs similar to bubbles of model 5.
The kinematics of the cooling gas (with temperature less than $10^7\,\rm{K}$ along the filament is displayed in
 
\begin{figure}
  \resizebox{\hsize}{!}{\includegraphics[angle=-90]{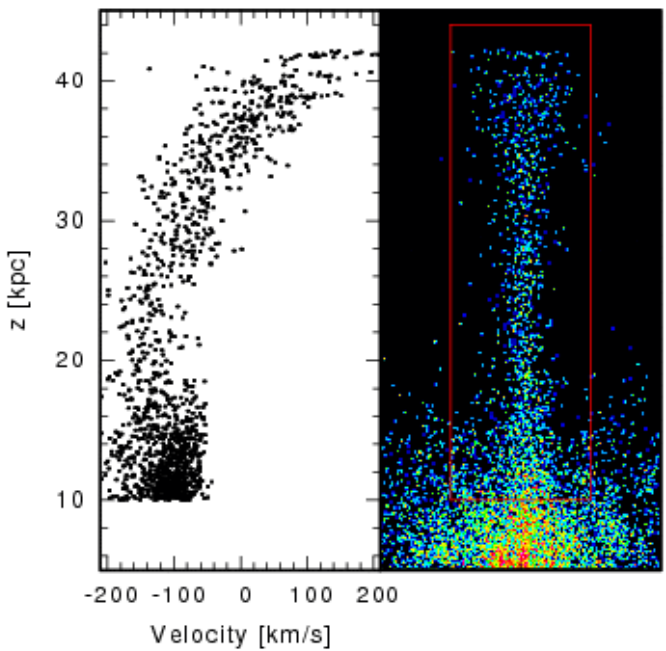}}
  \caption{Velocity of gas in the filament of model 5 at $t=400\,\rm{Myr}$ mesured along the $z$ axis. 
           \emph{left} : Distance of the filaments from the center, as a function of their speed.
           \emph{right} : Surface density of the gas. The particles represented in the left part of the figure are contained in
	   the red box.}
  \label{coldgas_speed}
  \end{figure}

Fig.~\ref{coldgas_speed}. The top of the filament is still entrained by the bubbles and has positive velocities
between $100$ and $200\,\rm{km/s}$, while the bottom of the filament is falling towards the center 
with negative velocities. This result, showing that the filaments are stretching, is qualitatively in 
agreement with the observations of \citet{hatch06}.
%


\section{Discussion and conclusion}\label{conclusion}


New very-high resolution SPH simulations of the Perseus cluster show
that the cold gas observed in H$\alpha$ and CO filaments outside the cluster core 
is a natural product of the perturbations of the ICM by buoyantly rising bubbles. 
During the rise of a bubble, a fraction of the low cooling time ambient ICM gas around the bubble 
is entrained and cools at higher radius, forming cold gas below $10^4\,\rm{K}$ in the trunk of the bubble.
The streaching of the filaments observed in H$\alpha$ by \citep{hatch06} is qualitatively reproduced.
%
%
%

Because no central heating is taken into account, all our models are subject to the cooling catastrophe.
Is the proposed scenario still valid in a model where the cooling flow is quenched by heating processes ?
The role of the heating source is to suppress the formation of gas with a temperature lower than $2\,\rm{keV}$,
but $1-2\,\rm{keV}$ gas will still be present in abundance in the cluster center \citep{fabian06}. 
As seen in section \ref{cold_trunk}, this is precisely this gas that is trapped by the bubbles and later forms 
the cold filaments. The heating at the origin of the cooling flow quenching will thus have only a very weak influence on the amount 
of cold gas formed. 
%
%
%
%
In the new emerging picture of cooling flow, AGN feedback provides some heating 
(negative feedback), but it is also responsible for a positive feedback, where ICM perturbations induce cold gas production. 
This is a step forward in the `cold feedback` scenario \citep{soker01,pizzolato05}
where the feedback takes place within a large region ($R\lesssim 5-30\,\rm{kpc}$).


\begin{acknowledgements}
      We are grateful to the anonymous referee for interesting and  helpful comments.
      This work has been supported by the Horizon project.
\end{acknowledgements}

\bibliographystyle{aa}
\bibliography{bibliography}

\Online

\begin{appendix}

\section{Control run without feedback}\label{appendix1}

\begin{figure}
  \resizebox{\hsize}{!}{\includegraphics[angle=-90]{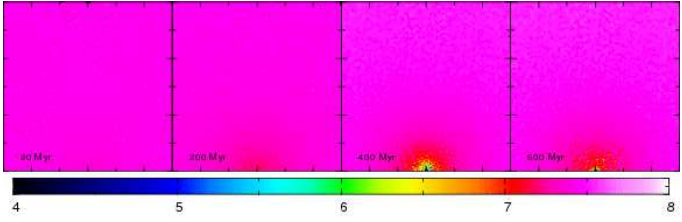}}
  \caption{Temperature map of the evolution of the cluster for the control simulation run with no bubbles. 
  Colors represent the logarithm of the temperature between $10^4$ and $10^8\,\rm{K}$ as indicated by the color bar.
  Cold gas is then represented by blue colors. The temperatures are computed from a thin slice of $2\,\rm{kpc}$
  width centered on $y=0$.
  The width and height of each box is $150\,\rm{kpc}$}
  \label{bubbles35}
  \end{figure}

\begin{figure}
  \resizebox{\hsize}{!}{\includegraphics[angle=0]{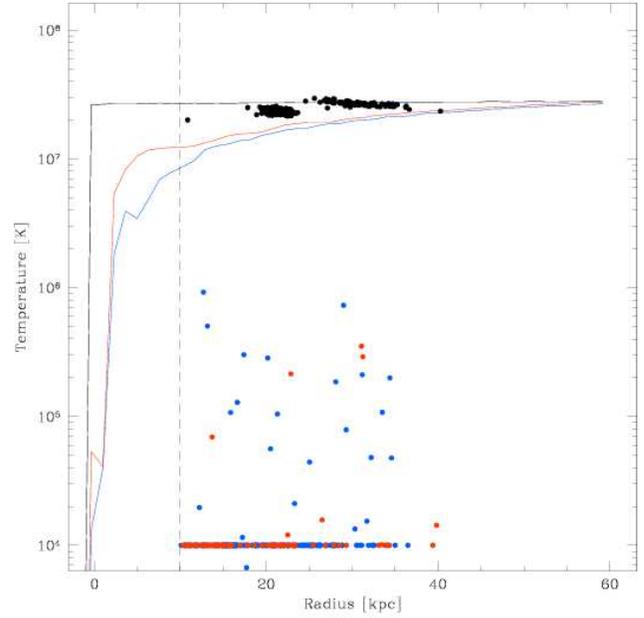}}
  \caption{Evolution of the mean temperature profile of the reference model, corresponding to a purely cooling flow 
  (black : $t=20\,\rm{Myr}$, blue : $t=400\,\rm{Myr}$, red : $t=600\,\rm{Myr}$). 
  The dashed vertical line corresponds to our definition of the core radius ($T<10^7\,\rm{K}$).
  Dots corresponds to the positions of particles forming a cold filament, at $t=400$ and $t=600\,\rm{Myr}$
  (black : $t=20\,\rm{Myr}$, blue : $t=400\,\rm{Myr}$, red : $t=600\,\rm{Myr}$, see also Fig.~\ref{bubble19_24_400}).}
  \label{Tr}
  \end{figure}

We present the result of the control simulation run without feedback.
Fig.~\ref{bubbles35} (the equivalent of Fig.~\ref{bubbles}) shows the temperature map
of the cluster. 
The cooling catastrophe occurs after $300\,\rm{Myr}$. After $600\,\rm{Myr}$, only the 
central $10\,\rm{kpc}$ have temperatures below $10^7\,\rm{K}$. Fig.~\ref{Tr} shows the evolution
of the mean temperature profile of the cluster. The cluster core is defined as the region
where the temperature drops below $10^7\,\rm{K}$.

\begin{figure}
  \resizebox{\hsize}{!}{\includegraphics[angle=0]{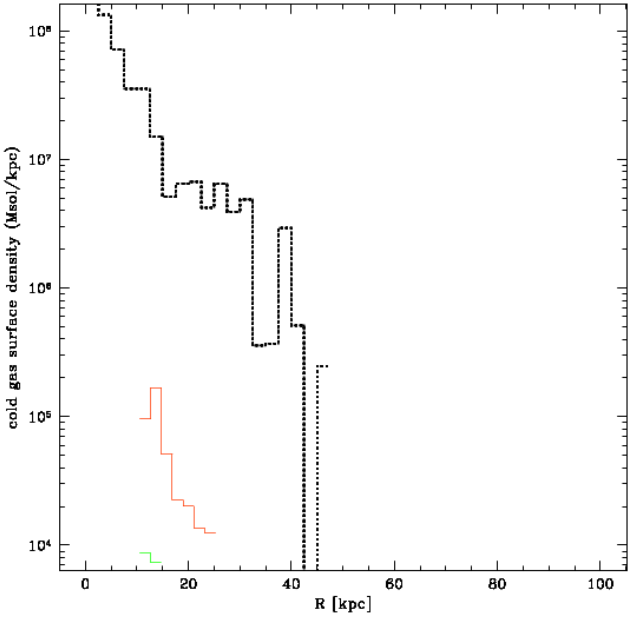}}
  \caption{Cold gas surface density as a function of radius for the control run,
  at $t=400$ (green) and $t=500\,\rm{Myr}$ (red). 
  The dashed line corresponds to values derived from \citet{salome06}.}
  \label{co40035}
  \end{figure}

Fig.~\ref{co40035} 
shows the cold gas surface density of the outlying regions ($R>10\,\rm{kpc}$).
Those values have been used to correct the equivalent values for models with bubble perturbation 
presented in in Appendix~\ref{appendix2}.

%
%

\section{Formation of cold filaments}\label{appendix0}

\begin{figure}
  \resizebox{\hsize}{!}{\includegraphics[angle=0]{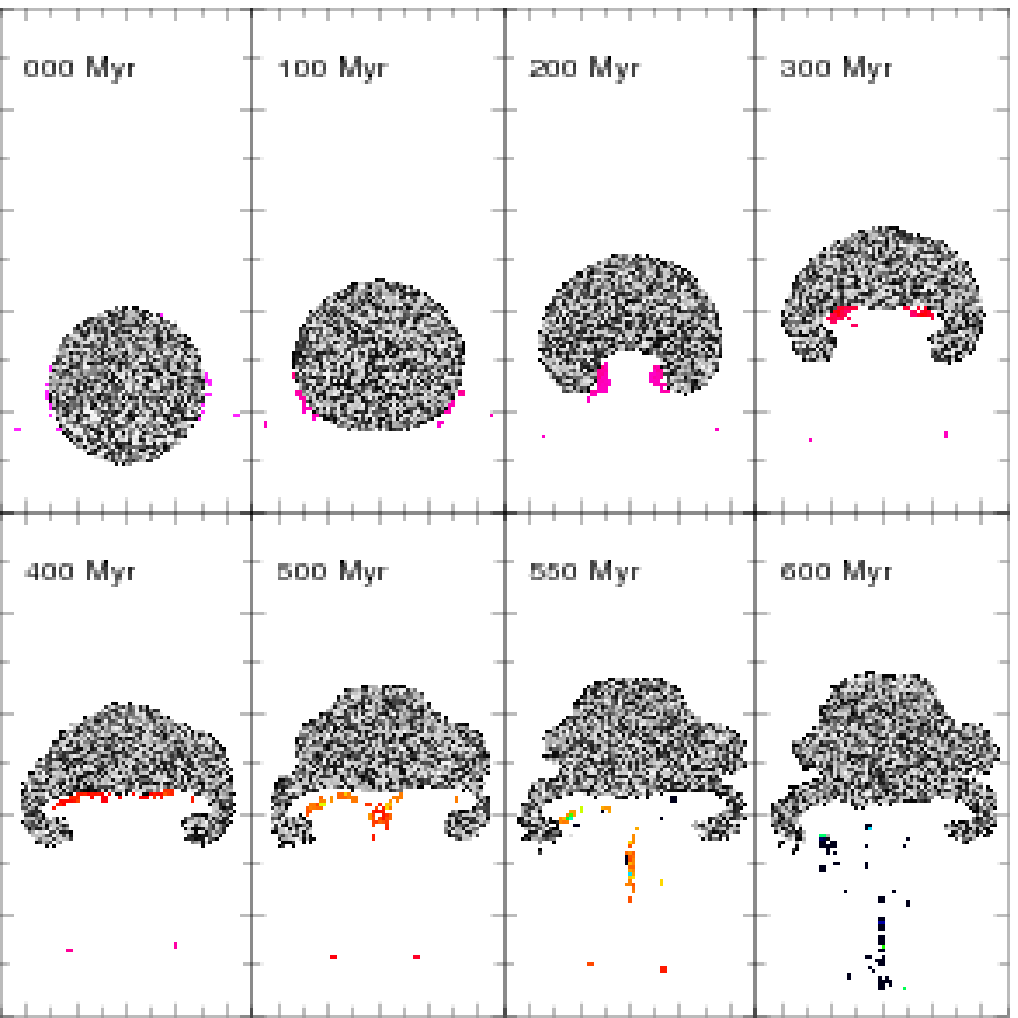}}
  \caption{Evolution of the gas around the bubble of model 2 that form the cold filament ($T<10^6\,\rm{K}$)
  at $t=600\,\rm{Myr}$. Colors represent the logarithm of the temperature between $10^4$ and $10^8\,\rm{K}$.
  The plasma bubble is represented in grey. The box size is $100\times200\,\rm{kpc}$.}
  \label{bubble19_600}
  \end{figure}

\begin{figure}
  \resizebox{\hsize}{!}{\includegraphics[angle=0]{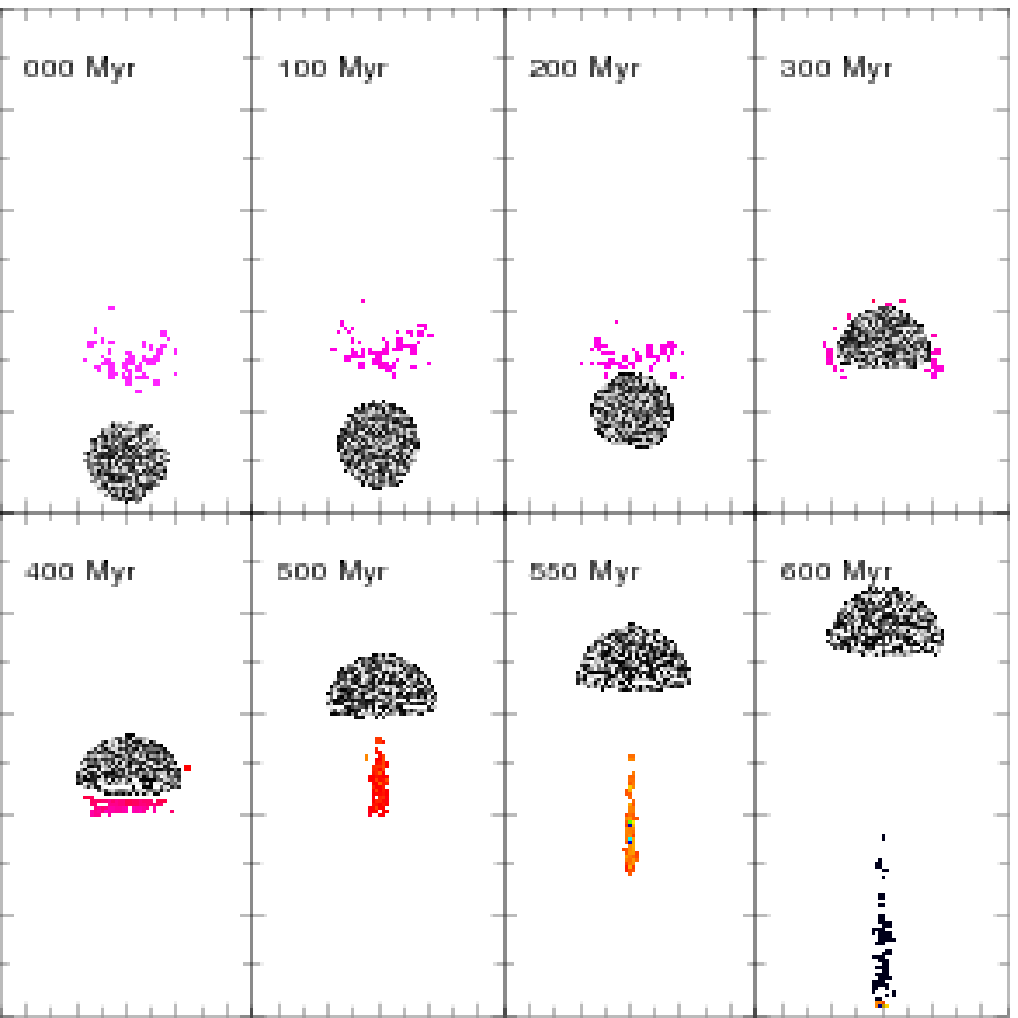}}
  \caption{Evolution of the gas around the bubble of model 4 that form the cold filament ($T<10^6\,\rm{K}$)
  at $t=600\,\rm{Myr}$. Colors represent the logarithm of the temperature between $10^4$ and $10^8\,\rm{K}$.
  The plasma bubble is represented in grey. The box size is $50\times100\,\rm{kpc}$.}
  \label{bubble24_600}
  \end{figure}

Fig.~\ref{bubble19_600} and \ref{bubble24_600} show the evolution and interaction with 
the rising bubble of the gas of model 2 and 5, that ends in the filament at $t=600$ with a 
temperature below $10^6\,\rm{K}$. 


\section{Comparison with CO observations}\label{appendix2}

In this appendix, we compute the amount of cold gas ($T<10^4\,\rm{K}$) formed outside the cluster core ($R > 10\,\rm{kpc}$), 
due to the uplift of the ambient ICM gas during the bubble rise and compare it with available CO observation of the Perseus cluster.
As no central heating is taken into account at the cluster center, a large amount of cold gas is naturally 
deposited within the central region (at a rate of about $160\,\rm{M_\odot/yr}$). From the amount of gas found for each simulation, 
we have then subtracted the equivalent value obtained from the control simulation (see Appendix~\ref{appendix1}), where no bubbles 
have been launched. This ensures that the cold gas observed for $R > 10\,\rm{kpc}$ is only due to ICM perturbation by the rising bubbles.

\begin{figure}
  \resizebox{\hsize}{!}{\includegraphics[angle=0]{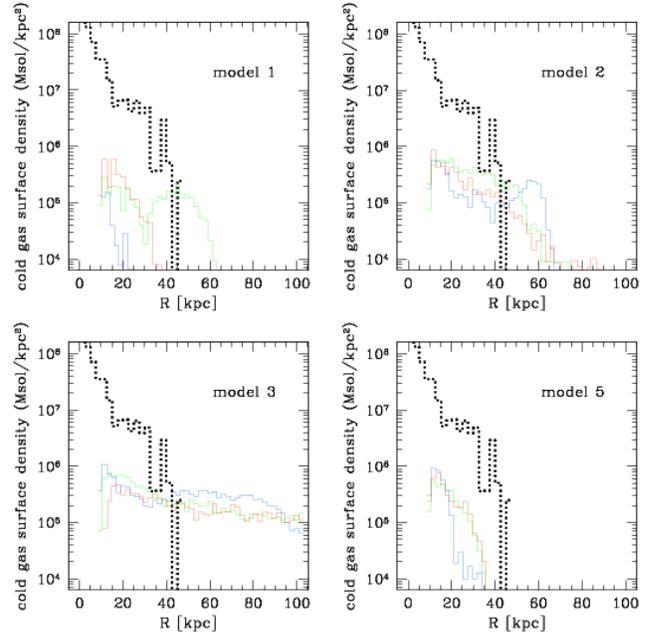}}
  \caption{Cold gas surface density as a function of radius for models 1, 2, 3, 5,
  at $t=300$ (blue), $t=400$ (green) and $t=500\,\rm{Myr}$ (red). 
  The dashed line corresponds to values derived from \citet{salome06}.}
  \label{co400}
  \end{figure}

Fig.~\ref{co400}, compares the cold gas surface density of models 1, 2, 3 and 5
\footnote{As the cold gas production of model 4 is nearly two decades below the observations, we haven't displayed it}, 
at three different times, with the surface density of the cold gas deduced from the \citet{salome06} data (dashed line). 
Clearly, all models fail to reproduce the amount of gas observed between $10$ and $40\,\rm{kpc}$.
However, we have to recall that in those models, only two symmetric bubbles are responsible for the ICM cooling outside the cluster core. 
In a more complete model, taking into account the recurrent or continuous activity
of the AGN will generate more bubbles, multiplying the production of cold gas by a dex.
It then means that models 1, 2 and 3 probably over predict the cold gas, especially at a radius larger than $40\,\rm{kpc}$. 
However, the surface density profile of model 5 at $t=400$ and $t=500$ becomes remarkably similar to the Perseus observations. 
This argumentation is supported by the direct comparison of cold gas in the HorseShoe region (Pos 11 in \citet{salome07}).
An amount of $0.9\cdot 10^8\,\rm{M_\odot}$ has been estimated in this region, remarkably similar to the cold gas formed in the 
rim of the bubbles of model 2, 3 and 4, ranging between $0.5$ and $10\times10^8\,\rm{M_\odot}$.

It is important to notice that a carefull comparison between the models and observations should also take
into account the lifetime of the filaments with respect of the AGN duty cycle. 
Unfortunately, the lifetime of the filaments is difficult to predict because it depends on several physical processes 
like heat conduction \citep{nipoti04}, viscosity or turbulence, not included in the present model.

\end{appendix}

\end{document}